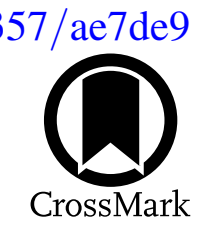

# Stellar Multiplicity in Open Clusters: Investigating Binary Fractions and Their Relationship with Cluster Properties

Ruan P. Alves, Hektor Monteiro, Wilton S Dias, and Gabriel R. Hickel
Instituto de Física e Química, Universidade Federal de Itajubá, Av. BPS 1303 Pinheirinho, 37500-903 Itajubá, MG, Brazil

## Abstract

Stellar multiplicity is key in dynamical evolution of open clusters (OCs) and processes of stellar formation and evolution. Since most stars form in multiple systems, investigating how environmental properties affect their occurrence is crucial. We aim to investigate how binary fraction and mass ratios change across different OC environments. Using a sample of 773 OCs, we identified binary systems and estimated binary fractions, obtaining values in the ranges 0.10–0.85 overall and 0.06–0.79 for high mass-ratio systems. No strong dependence on age or evolutionary stage is found, while an anticorrelation between metallicity and the binary fraction is identified. However, young and old clusters exhibit distinct mass-ratio distributions. We also find that the probability of a star being the primary component increases with stellar mass. In contrast, the probability of a star belonging to a binary system increases with mass for $M < 1\,M_\odot$ and remains between 60% and 70% for high masses. Binary systems are preferentially located in the inner regions of clusters compared to single stars, and the mean mass ratio decreases with increasing distance from the cluster center. These results suggest that the binary fraction appears to be set at early times and remains largely unaffected by cluster dynamical evolution. The concentration of binaries in the inner regions of clusters likely reduces their susceptibility to dynamical escape.

## 1. Introduction

Binary or multiple systems are common products of the star formation process (G. Duchêne & A. Kraus 2013; H. M. J. Boffin & D. Jones 2025), with most solar-type stars having a companion (K. Fuhrmann et al. 2017) and practically all high-mass stars being part of a multiple system (H. Sana et al. 2012; R. Chini et al. 2013). These objects have significant effects on the dynamics of stellar clusters (R. de Grijs et al. 2008; S. Rastello et al. 2020) and planetary formation (W. Kley & R. P. Nelson 2008; K. Silsbee & R. R. Rafikov 2021; J. E. Hand et al. 2025), while evolved binaries can give rise to exotic objects such as supernova explosions (Z.-W. Liu et al. 2023), X-ray binaries (M. Falanga et al. 2021), and, in some cases, a detectable source of gravitational waves (T. M. Tauris & E. P. J. van den Heuvel 2023).

One possible environment for studying stellar multiplicity is open clusters (OCs). Composed of tens to thousands of stars formed from the same molecular cloud, we can assume that the member stars of an OC share the same fundamental properties, such as age, chemical composition, distance, and reddening. In this way, it is possible to analyze the stellar populations of different clusters and obtain important insights into stellar formation and evolution in diverse environments.

Although identifying binary systems is a complex observational problem, particularly for close and wide binaries, many unresolved systems can be identified through the cluster color–magnitude diagram (CMD). Compared to single stars of a given mass, binary systems exhibit photometry shifted toward brighter and redder regions of the CMD due to the contribution of the companion star. The magnitude of this shift increases with the mass ratio $q$[1] of the system.

Studies of OC properties, including their binary fraction ($f_b$), have a long history in the literature. For example, J.-C. Mermilliod & M. Mayor (1989) reported binary fractions ranging from 22% to 35% for five OCs, while E. Bica & C. Bonatto (2005) and S. Sharma et al. (2008) derived values of 15%–54% and 30%–40%, respectively, each analyzing samples of nine OCs. The high-quality observations provided by the Gaia mission (Gaia Collaboration et al. 2023b) have enabled significant advances in the identification of unresolved binaries, and the release of the Non-Single Stars catalog (Gaia Collaboration et al. 2023a) has provided a large sample for several studies (e.g., T. Jayasinghe et al. 2023; Z. D. Hartman et al. 2025; A. Mazzi et al. 2025), generating a considerable advance in the study of stellar multiplicity.

More recent studies have explored the binary fraction in OCs using a variety of approaches. Using synthetic CMDs, H. Niu et al. (2020) estimated binary fractions in 12 OCs, finding values between 29% and 55%. Meanwhile, using Gaia eDR3 data, Y. Jiang et al. (2024) estimated the fraction of binaries with mass ratios $q > 0.5$ in 114 clusters within 500 pc, obtaining values between 6% and 34%. Similarly, J. Donada et al. (2023) and V. V. Jadhav et al. (2021) analyzed stellar multiplicity in 202 and 23 OCs, respectively, considering systems with $q > 0.6$, and reported binary fractions ranging from 5% to 67% and 12% to 38%. Treating the observed CMD as a mixture of single stars, unresolved binaries, and field stars, L. Li & Z. Shao (2022) estimated $f_b$ for systems with $q > 0.2$ in 10 OCs, obtaining values between 30% and 50%.

[1] $q = M_2/M_1$, where $M_2$ and $M_1$ are the companion and primary masses, respectively ($M_2 \leqslant M_1$).

A. Almeida et al. (2023), using a new method to estimate individual stellar masses, calculated the total mass of a large sample of OCs, containing a total of 773 clusters from the W. S. Dias et al. (2021) catalog. The method is based on the comparison of observed data from Gaia eDR3 with synthetic clusters, with masses estimated by the Monte Carlo method. In addition to providing accurate estimates of stellar masses, this approach also allows for the identification of potential candidate binary systems, which are adopted in this study.

In this work, we use a large sample of OCs provided by A. Almeida et al. (2023) to investigate how fundamental cluster parameters, such as age and metallicity, influence the binary fraction in OCs. We also examine how stars in different mass ranges participate in binary systems. In addition, we analyze the spatial distribution of binaries within clusters and discuss their implications for the dynamical evolution of these environments.

This paper is organized as follows. In Section 2, we describe the cluster sample analyzed in this work. The methodology is presented in Section 3. The results for the binary fraction and the mass ratios are presented in Section 4. In Section 5, we compare our estimates for the binary fraction with the values found in the literature. The relationship between the binary fraction and the properties of clusters is discussed in Section 6. Finally, our conclusions are presented in Section 7.

## 2. Cluster Sample

In this work, we used the sample of OCs published by A. Almeida et al. (2023). For the stellar membership probabilities and fundamental parameters of the clusters, we adopted the same values published by W. S. Dias et al. (2021), while the photometric and astrometric data were updated with Gaia DR3 data (Gaia Collaboration et al. 2021; European Space Agency & Gaia Collaboration 2022).

The clusters span distances of 0.14–6.87 kpc, extinctions of $A_V = 0.01$–4.49 mag, ages from 4 Myr to 6 Gyr, and metallicities ([Fe/H]) between −0.37 and 0.43 dex. These homogeneous data and broad parameter space, as presented in Figure 1, allow a detailed statistical analysis of stellar multiplicity under diverse Galactic conditions.

## 3. Methodology

### 3.1. Stellar Masses and Selection of Binary Systems

In this study, we used directly the results obtained by our team as published in A. Almeida et al. (2023). The procedures used to estimate stellar masses and identify binary systems are briefly presented below, and all details can be found in A. Almeida et al. (2023).

For each cluster, the fundamental parameters (age, metallicity, distance, and extinction) and the photometric data were used to generate synthetic clusters. For each observed star, a sample of masses was generated via a Monte Carlo method, and based on comparisons with the synthetic clusters, each member star was classified as single or binary, allowing the estimation of the mass of the main star as well as that of its companion. The final stellar masses were obtained by averaging the Monte Carlo realizations, and their uncertainties were computed as the standard deviation of the mass distribution.

The synthetic clusters were constructed using the Padova PARSEC (version 1.2S) set of theoretical isochrones, described by A. Bressan et al. (2012). Each synthetic cluster contains 10,000 stars and assumes a binary fraction of 50%. Stellar masses were drawn from a probability distribution described by the Chabrier initial mass function (G. Chabrier 2003), while companion masses were sampled from a uniform distribution between zero and the mass of the primary star. The procedure consider both observational and synthetic photometric errors. In this work, we exclude systems identified as binaries for which $\delta M_2/M_2 \geqslant 0.5$, where $M_2$ is the companion mass and $\delta M_2$ is its associated uncertainty.

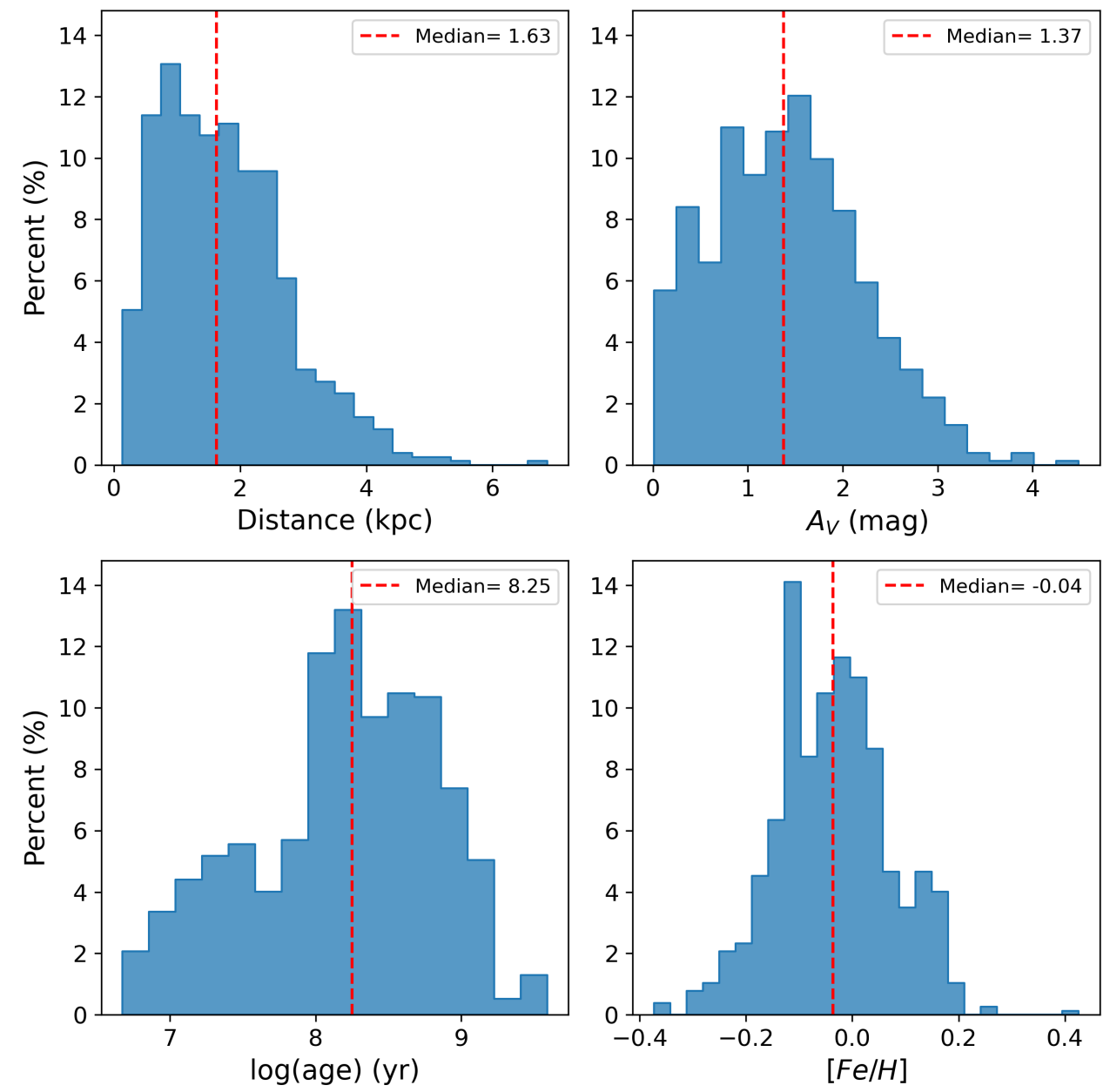

**Figure 1.** Distribution of the parameters of the sample of 773 OCs used in this study. The dashed lines indicate the median of each parameter.

The binary fraction $f_b$ of each cluster was determined using the following definition:

$$f_b = \frac{N_b}{N_s + N_b}, \quad (1)$$

where $N_b$ and $N_s$ are the numbers of binary and single systems in the cluster, respectively. We also estimated the high mass-ratio binary fraction ($f_b^{q>0.5}$), considering only systems with $q > 0.5$ as binary.

### 3.2. Sample Bias

In our sample, we identify a systematic bias in the estimated binary fraction associated with distance and reddening effects, as shown in Figure 2. The distributions of $f_b$ exhibit a clear shift toward higher values for distant clusters (left panel) and for clusters affected by higher extinction $A_V$ (right panel).

At larger distances, the observed stellar population is increasingly biased toward higher-mass stars due to the limited detectability of faint sources. Since massive stars are preferentially found in multiple systems (H. Sana & C. J. Evans 2011), the binary fraction tends to be artificially increased in distant clusters. On the other hand, extinction $A_V$ shifts stellar photometry toward redder colors in the CMD, while differential reddening may broaden the main sequence by introducing additional scattering in stellar colors and magnitudes. Because the method adopted in A. Almeida et al. (2023) identifies unresolved binaries based on photometric deviations relative to single-star sequences, both the reddening-induced displacement of the CMD and the additional broadening caused by differential reddening may increase the likelihood of falsely classifying stars as binaries, potentially leading to an overestimation of the binary fraction in highly reddened clusters. However, differential reddening was not explicitly taken into account in the present study.

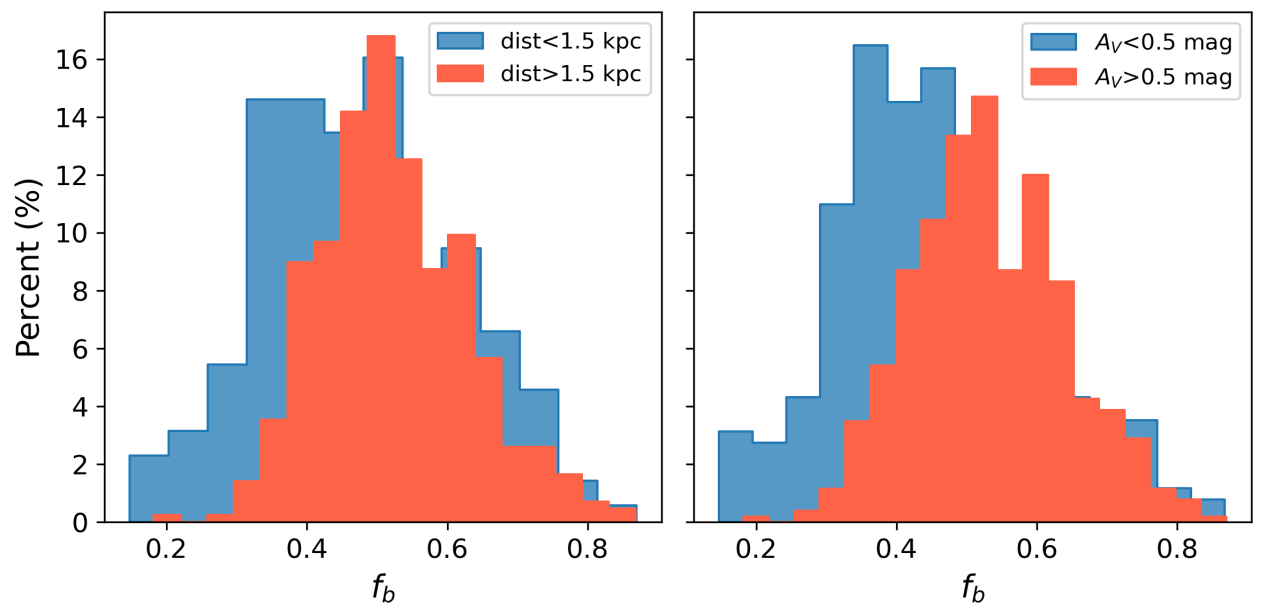

**Figure 2.** Distributions of the binary fraction for subsamples of OCs separated by distance (left panel) and extinction $A_V$ (right panel).

### 3.3. Correction of the Binary Fraction

To estimate a correction for binary fractions of the OCs associated with distance and reddening effects, we started from the hypothesis that, on average, the $f_b$ should not vary significantly with distance or reddening and that these properties are mainly related to observational issues rather than to the intrinsic properties of the binary systems themselves.

Based on this premise, we defined a benchmark subsample composed of 96 OCs with distances smaller than 1.5 kpc and reddening smaller than 0.5. Within this range of parameters, the values of $f_b$ can be considered to be more reliable, given the lower impact of observational uncertainties and completeness. The description of the OCs properties in the benchmark sample, as well as a comparison with the clusters in the full sample, is shown in the Appendix A.

With the benchmark sample defined, we initially applied three independent approaches to correct the binary fraction: (i) a correction based on mean values in the benchmark subsample, (ii) a moving-average (MA) correction, and (iii) a similarity-based correction.

In the first approach, we computed the mean binary fraction of the benchmark subsample, $\langle f_b \rangle_{\rm bench}$. The full sample was then divided into bins of $A_V$ with a width of 0.5 mag. For each bin, we calculated the mean binary fraction $\langle f_b(A_V) \rangle$ and defined a correction term as

$$\delta = \langle f_b(A_V) \rangle - \langle f_b \rangle_{\rm bench},$$

and the corrected binary fraction for each bin of $A_V$ was then obtained as $f_b^{\rm corr,mean} = f_b - \delta$.

In the second approach, clusters were first sorted by $A_V$, and an MA smoothing was applied to the corresponding $f_b$ values along this ordered sequence. For each cluster $i$, the MA was computed over a centered local window of neighboring clusters in the ordered $A_V$ sequence as

$$MA_i = \frac{1}{w} \sum_{k=i-w/2}^{i+w/2} f_{b,k}, \quad (2)$$

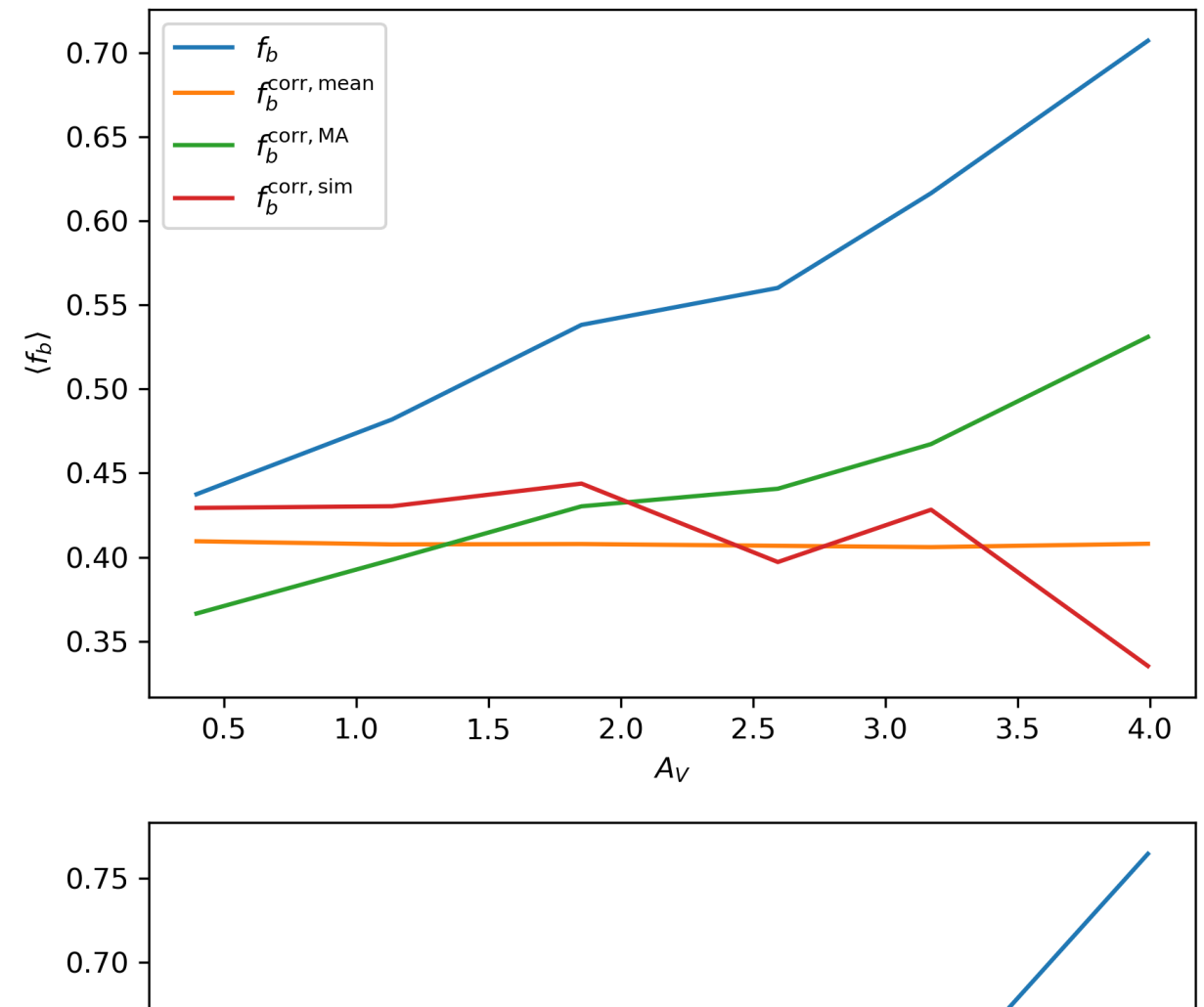

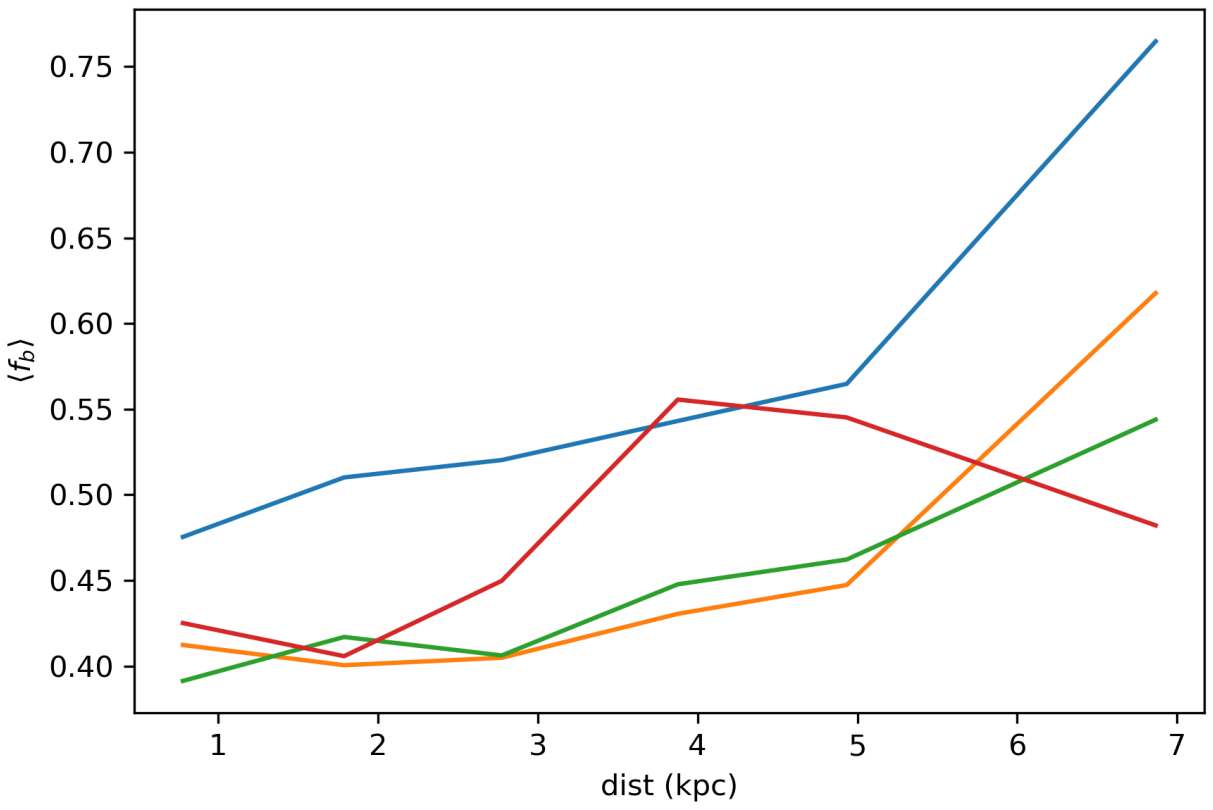

**Figure 3.** Binary fraction as a function of $A_V$ (top panel) and distance (lower panel). The blue line shows the original values, while the orange, green, and red lines correspond to the mean, MA, and similarity-based corrected $f_b$, respectively.

where $w$ is the window size. We adopted a centered MA with $w = 5$ clusters. The corrected binary fraction was then defined as $f_b^{\rm corr,MA} = f_b - (MA_i - \langle f_b \rangle_{\rm bench})$.

Finally, in the third approach, we estimated the corrected binary fraction based on the similarity between clusters. For each cluster in the full sample, we identify the five closest clusters in age, metallicity, number of members, and total mass within the benchmark subsample. Using these neighbors, we define the corrected values for the binary fraction as a weighted mean:

$$f_b^{\rm corr,sim} = \frac{\sum_{n=1}^{5} w_n f_{b,n}}{\sum_{n=1}^{5} w_n}, \quad (3)$$

where $f_{b,n}$ is the binary fraction of the $n$th nearest cluster in the benchmark subsample and $w_n$ is a weight defined as the inverse of the distance $D_n$ in parameter space. The distance between two clusters $i$ and $j$ in this parameter space was calculated using the Euclidean distance:

$$D_{i,j} = \sqrt{\sum_{k=1}^{n} (P_k^{(i)} - P_k^{(j)})^2}, \quad (4)$$

where $P_k^{(i)}$ and $P_k^{(j)}$ are the values of the parameters $k$ for clusters $i$ and $j$, respectively. All parameters were expressed in linear space and normalized before to computing the distances between clusters.

Figure 3 shows the effects of $A_V$ and distance on the $f_b$ values for all applied corrections. As shown, the mean-based correction is the only approach that completely removes the dependence of the $f_b$ on $A_V$. The MA correction only smooths the reddening effects, while the similarity-based correction presents more stable values across the entire range of $A_V$. Despite the reduced effect of distance on the binary fraction, the trend remains unchanged regardless of the adopted correction. Based on the behavior of the different methods, we conclude that the mean-based correction is the most effective in mitigating the observational effects present in our sample. Therefore, we adopt the mean-based correction as our reference value for the binary fraction and apply the same correction to $f_b^{q>0.5}$. Hereafter, all results presented in this work consider the corrected values of $f_b$.

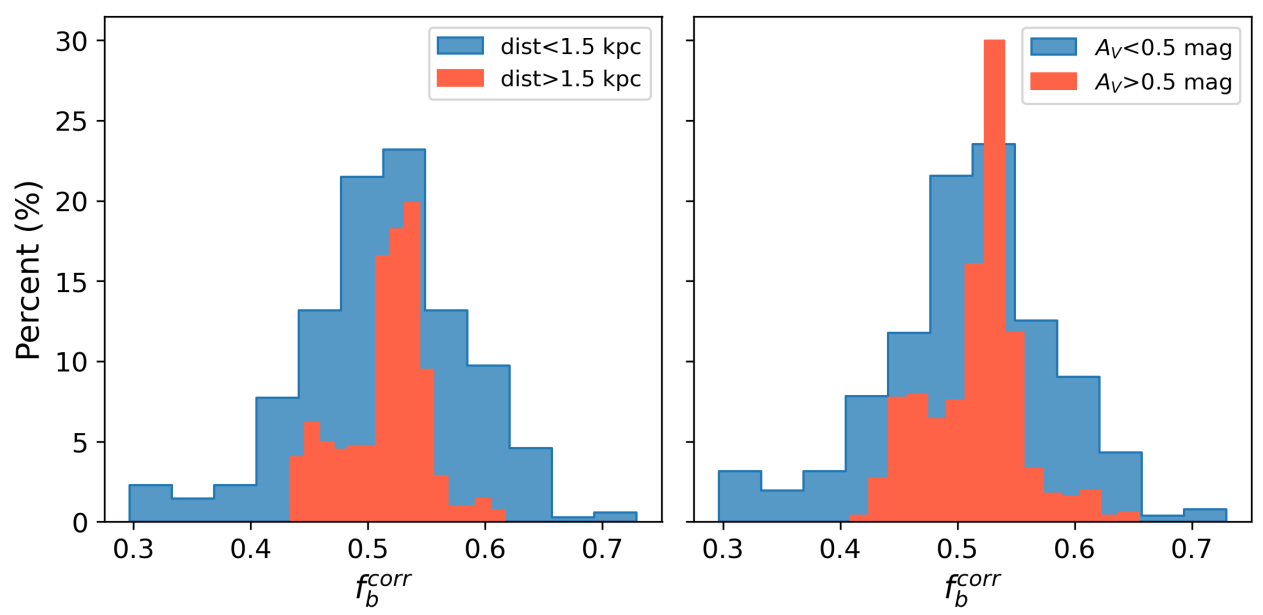

**Figure 4.** Distributions of the corrected binary fraction for subsamples of OCs separated by distance (left panel) and extinction $A_V$ (right panel).

In Figure 4 are given the same plots of Figure 2 but they are corrected by the distance and reddening effects. It shows reduced dispersion for distant and high reddening clusters, while no clear systematic shift is apparent between the distributions as a function of cluster distance or visual extinction $A_V$.

### 3.4. Uncertainty in the Binary Fraction

Taking into account the systematic trends discussed in Section 3.2, we estimated the uncertainties in both $f_b$ and $f_b^{q>0.5}$ by combining systematic ($\sigma_{\rm sys}$) and statistical ($\sigma_{\rm stat}$) uncertainties as shown below.

For each cluster, we first use the corrected value $f_b^{\rm corr}$ determined in Section 3.3 to estimate the number $N$ of systems potentially misclassified as single or binary as follows:

$$N = (f_b - f_b^{\rm corr})\, N_{\rm systems}, \tag{5}$$

where $N_{\rm systems}$ is the total number of cluster's stellar systems.

Next, we employed a Monte Carlo resampling procedure, using 1000 executions for each cluster. In each realization, we selected a random integer $N_{\rm boot}$ between 0 and $|N|$, where $|N|$ is the absolute number of systems to be reassigned. If $N > 0$, we randomly selected $N_{\rm boot}$ systems originally classified as binaries and reassigned them as single systems. Similarly, if $N < 0$, $N_{\rm boot}$ systems originally classified as single were reassigned as binaries. The binary fraction was then recomputed for each execution, and the systematic uncertainty $\sigma_{\rm sys}$ was defined as the standard deviation of the binary fraction distribution obtained from the 1000 resamplings.

Assuming binomial statistics for the number of binary systems, the statistical uncertainty is given by

$$\sigma_{\rm stat} = \sqrt{\frac{f_b(1-f_b)}{N_{\rm members}}}\,, \tag{6}$$

and the final uncertainty in the binary fraction was estimated by $\sigma_{f_b} = \sqrt{\sigma_{\rm sys}^2 + \sigma_{\rm stat}^2}$.

### 3.5. Probabilistic Definitions of Binary Systems

In this work, we also define the probabilities of stars within a given mass range are found in a binary system ($P_{\rm BS}$) and as primary ($P_{\rm Prim}$) or secondary ($P_{\rm Comp}$) components of the pair.

For a given sample, the total number of stars is

$$N_\star = N_{\rm Singles} + N_{\rm Prim} + N_{\rm Comp},$$

where $N_{\rm Singles}$ denotes the number of single stars, $N_{\rm Prim}$ the number of stars that appear as primary components in binary systems, and $N_{\rm Comp}$ the number of stars that appear as secondary (companion) components within the same mass range.

Using this definition, the probabilities are given by

$$P_{\rm BS} = \frac{N_{\rm Prim} + N_{\rm Comp}}{N_\star}, \tag{7}$$

$$P_{\rm Prim} = \frac{N_{\rm Prim}}{N_\star}, \tag{8}$$

and

$$P_{\rm Comp} = \frac{N_{\rm Comp}}{N_\star}. \tag{9}$$

The uncertainties in the probabilities were estimated using 1000 bootstrap resamplings of the stellar masses. In each resampling, the stellar masses were randomly drawn from Gaussian distributions centered on their nominal values, with standard deviations given by their respective uncertainties.

### 3.6. Binarity and Cluster Dynamical State

To investigate the radial distribution of binaries within OCs and their effect on the mass segregation, we estimated, for each member star, its distance from the cluster center, and also estimated the half-mass–radius ($R_h$) for each cluster. The uncertainties of $R_h$ were estimated using the bootstrap method with replacement, with 1000 resamplings of the original sample from each cluster. For each resampling, the corresponding $R_h$ value was recalculated, and the uncertainty was defined as the standard deviation of the bootstrap distribution of the $R_h$ values. To compare the radial distribution of binary systems for the different clusters in our sample, we adopted the dimensionless $R/R_h$.

For the total cluster masses we adopted the values provided by A. Almeida et al. (2023). We note that these mass estimates rely on the extrapolation of the present-day mass function and therefore assume that the underlying stellar population is not significantly affected by dynamical of the OC. However, some clusters may exhibit truncation of the mass function due to the preferential evaporation of low-mass stars. Additionally, the adopted spatial extent of each cluster may influence the inclusion of stars in the outer regions, potentially retaining evaporated members. These effects are not explicitly accounted for in this work; however, as our analysis focuses primarily on relative trends, these uncertainties are not expected to significantly affect our main results.

To further explore the role of dynamics in shaping binary populations, we analyzed the fraction of binaries as a function of the cluster's dynamical state, quantified by the parameter $\tau = {\rm age}/t_{\rm relax}$, where $t_{\rm relax}$ is the relaxation time of the cluster,

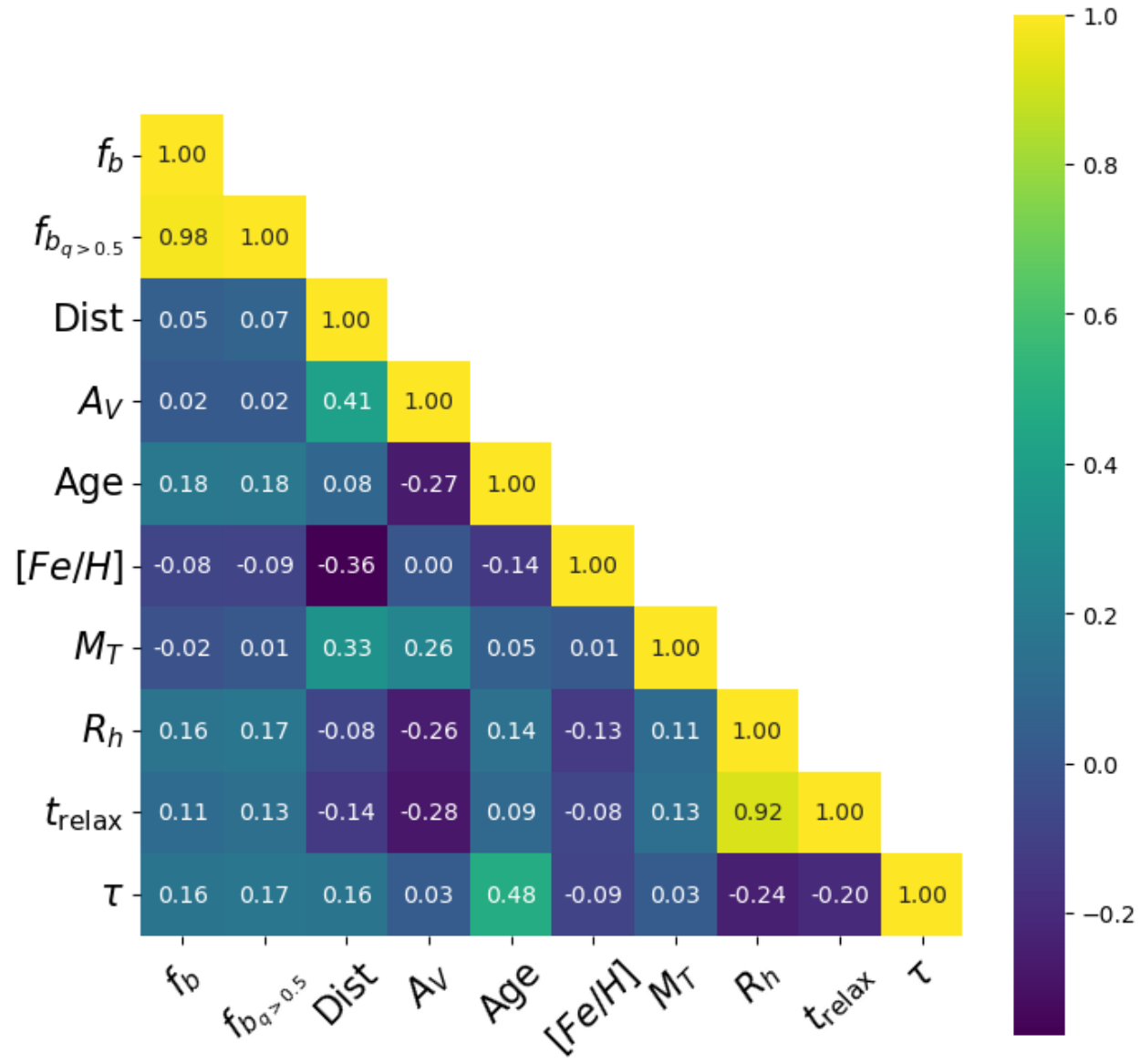

**Figure 5.** Confusion matrix between OCs parameters.

estimated following L. Spitzer & M. H. Hart (1971):

$$t_{\rm relax} = \frac{8.9 \times 10^5}{\log(0.4 N_{\rm memb})} \sqrt{\frac{N_{\rm memb} . R_h^3}{\bar{m}}}, \tag{10}$$

with $N_{\rm memb}$ being the number of cluster members and $\bar{m}$ its average stellar mass. The uncertainties in $t_{\rm relax}$ were estimated through standard error propagation, taking into account the uncertainties in $R_h$, and $\bar{m}$. The uncertainty in the dynamical parameter $\tau$ was then obtained by propagating the uncertainties in both age and $t_{\rm relax}$.

The Jupyter Notebook to execute the analysis in this paper is hosted on Zenodo (doi:10.5281/zenodo.20140202) and GitHub.[2]

## 4. Results

For each cluster, the catalog includes the estimated values of $f_b$, $f_b^{q>0.5}$, the half-mass ratio, the relaxation time, and the dynamical state parameter $\tau$, along with their associated uncertainties. Table 3 in Appendix B presents a preview of the catalog contents.[3] Figure 5 shows the confusion matrix among the cluster parameters, where we adopt Pearson's correlation coefficient ($r$) to quantify the degree of linear association between pairs of variables.

In the context of Figure 5, the metallicity, total mass, and cluster age do not exhibit strong correlations with the total binary fraction, with correlation coefficients of $-0.08$, $-0.02$, and 0.18, respectively. We also find no significant correlation between the total binary fraction and either the half-mass ratio or the dynamical state parameter. Similar behavior is observed in the correlations between the cluster parameters and $f_b^{q>0.5}$.

### 4.1. Mass Ratios

The values for the mass ratio of the binaries are distributed between 0.34 and 0.99, with a median of $q = 0.76$, suggesting a preference for high-$q$ systems. In our results, we observe a relative excess of nearly equal-mass binaries ($q \sim 0.9$), similar to those reported by, e.g., G. Duchêne & A. Kraus (2013) and H. Niu et al. (2020). Figure 6 shows the distribution of the derived mass ratios for all binary systems in the sample.

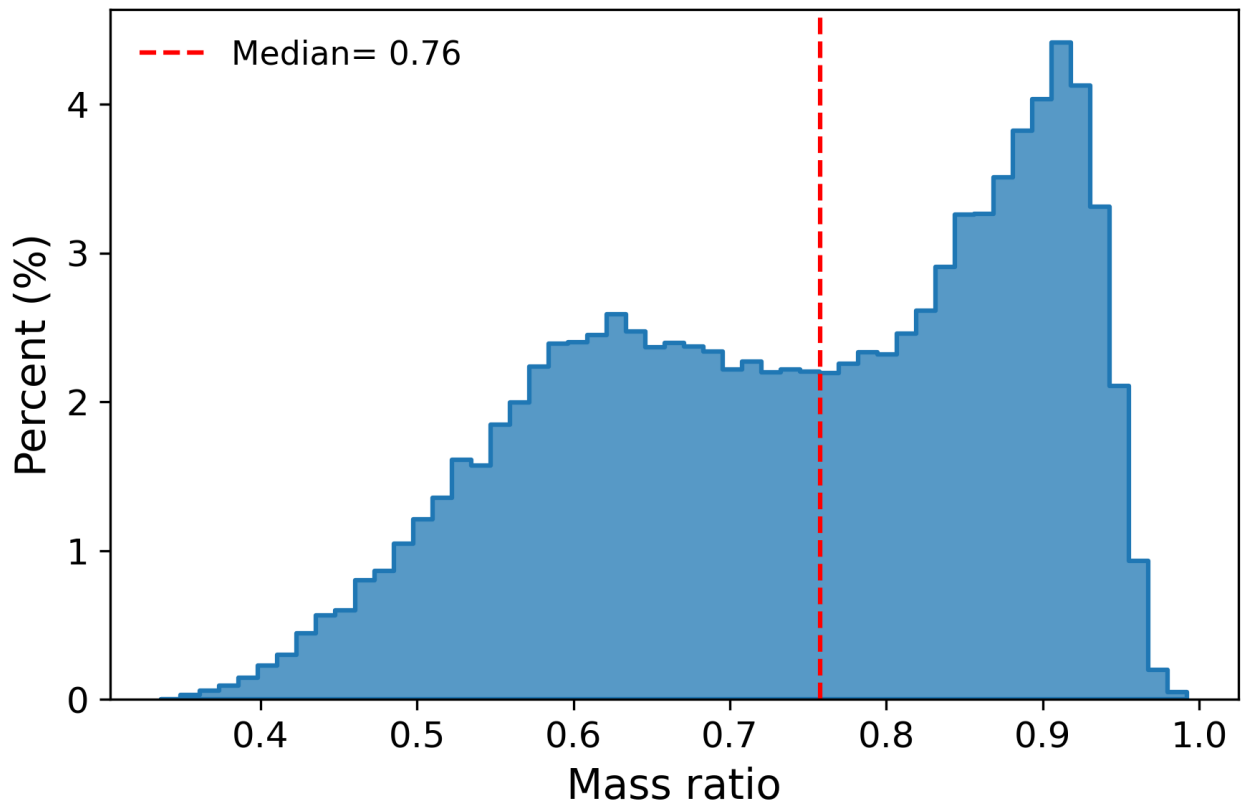

**Figure 6.** Distribution of mass ratios of all binary systems analyzed in this study.

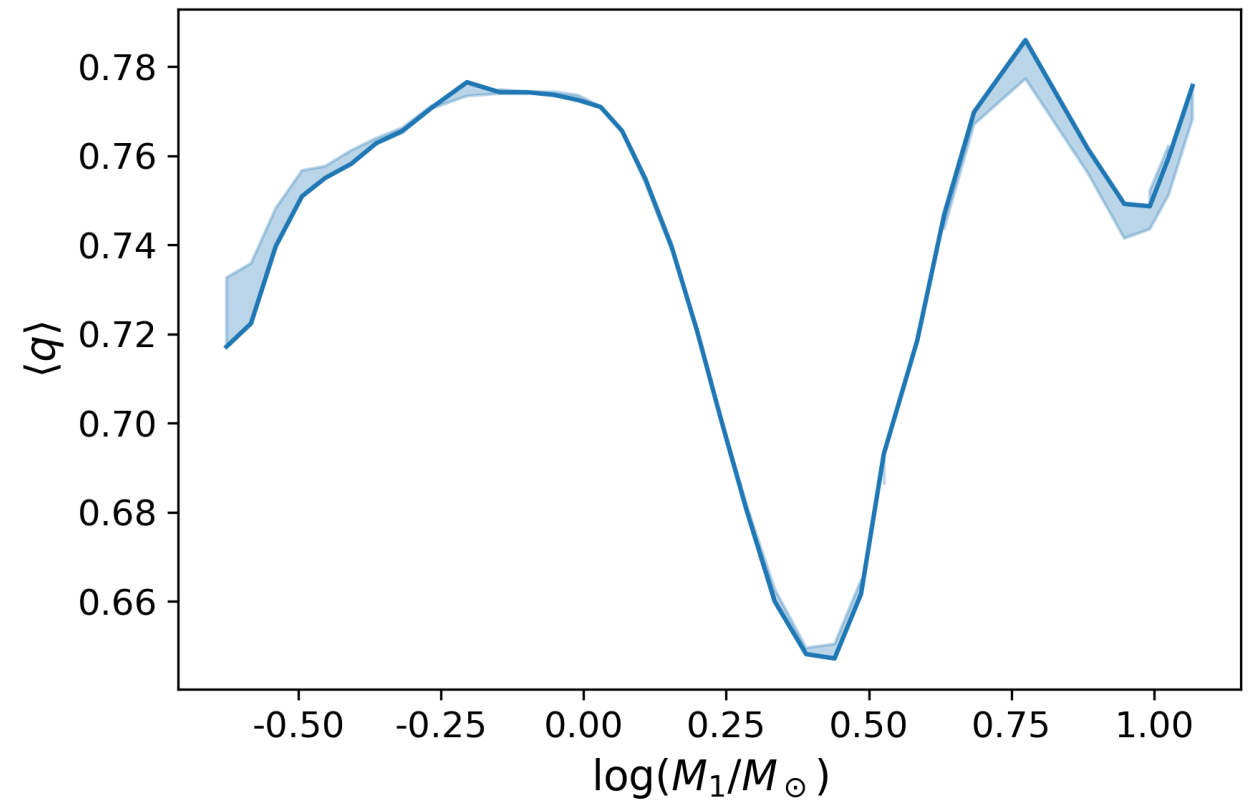

**Figure 7.** Mean mass ratio as a function of primary mass. The blue shaded region represents the uncertainty range.

We also find that the average mass ratio presents similar values ($\langle q \rangle \sim 0.75$) for primaries with masses less than $1\,M_\odot$ as well as for masses greater than $5\,M_\odot$. However, we report a significant decrease in $\langle q \rangle$ in function of the primary mass within the range of 1–3 $M_\odot$, with a minimum of $\langle q \rangle = 0.65$ in $M_1 \sim 2.75\,M_\odot$, as shown in Figure 7. The uncertainty band presented in the figure was estimated following the same procedure described in Section 3.5 for the calculation of the probability uncertainties.

We verified that the decrease in $\langle q \rangle$ is not restricted to the youngest clusters ($\log(\text{age}) < 7.5$) but is also present in intermediate-age clusters ($8.0 < \log(\text{age}) < 9.0$), suggesting that it is unlikely to arise exclusively from stellar evolutionary effects associated with very young populations. We also visually inspected the CMDs of representative clusters contributing to this mass interval and found that stars with $2 < M_1 < 3$, $M_\odot$ are generally located near the upper main sequence and turnoff region, without evidence of a clear or systematic population of evolved stars capable of explaining the observed dip.

Stellar rotation effects could also contribute to this behavior, since rapid rotation may broaden the main sequence and potentially lead to the misclassification of single stars as binaries within the adopted method. However, rotation effects were not considered in the present work, and with the available data we are unable to directly assess whether this mechanism contributes significantly to the observed feature.

[2] https://github.com/RuanPH26/stellar_multiplicity_in_open_clusters

[3] The complete results of this work are provided in an online table.

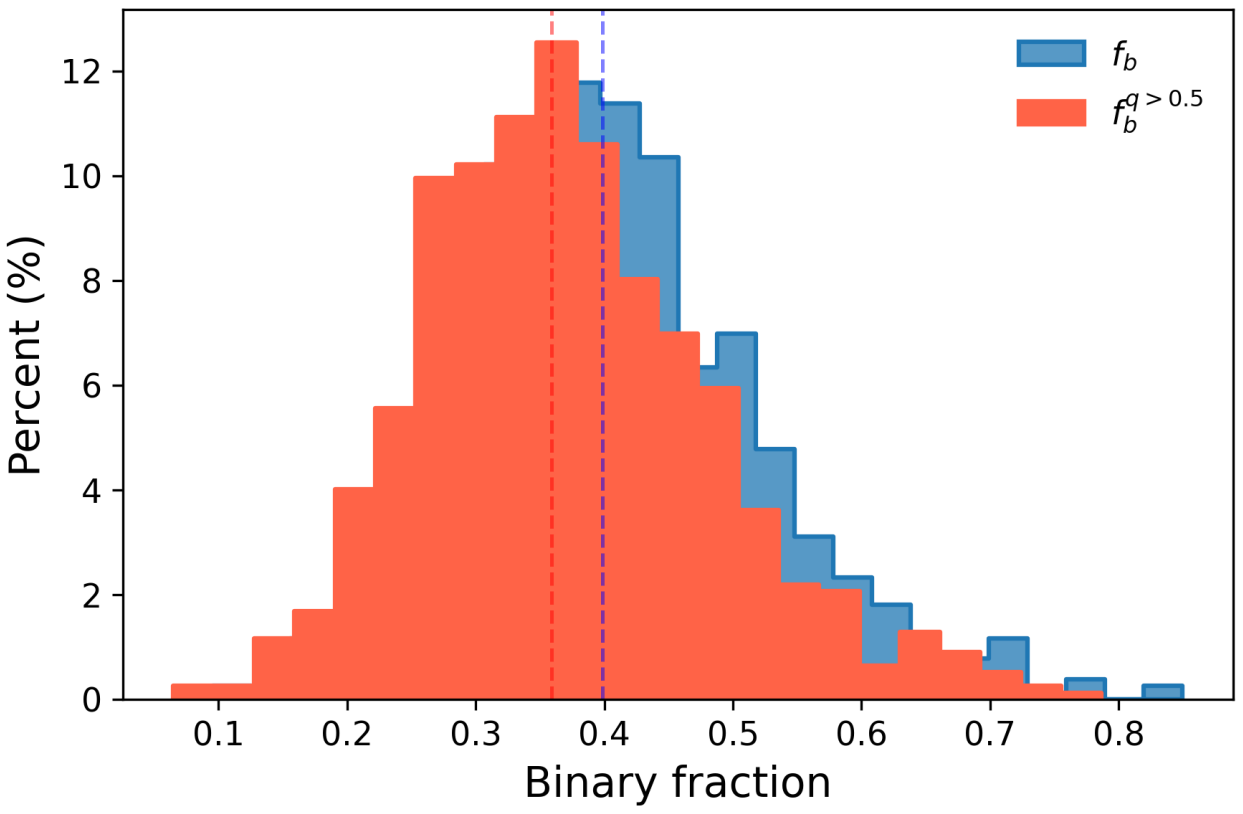

**Figure 8.** Distribution of the values of $f_b$ (blue) and $f_b^{q>0.5}$ (red). Dashed lines indicate the median values, 0.40 and 0.36, respectively.

### 4.2. Binary Fraction

The total binary fraction estimated for the OCs in our sample ranges from 0.10 to 0.85, with a median value of 0.40. For high mass-ratio systems, the binary fraction is lower, with $f_b^{q>0.5}$ ranging from 0.06 to 0.79 and a median of 0.36.

The distributions of $f_b$ and $f_b^{q>0.5}$ are shown in Figure 8. Although their overall shapes are similar, the $f_b$ distribution is slightly shifted toward higher values. The mean uncertainties in $f_b$ and $f_b^{q>0.5}$ are both 0.15, with maximum values of 0.41 and 0.38, respectively.

It is important to note that our estimates of the binary fraction are restricted to systems with $q > 0.34$. As a result, binaries with low-mass companions are not fully accounted for in our analysis. Therefore, the values of $f_b$ reported here should be interpreted as lower limits to the true binary fraction in OCs.

## 5. Comparison with the Literature

The methods for determining binary systems, as well as the criteria for selecting the systems considered, vary widely in the literature. We compared our estimates of $f_b$ with values reported in the works of J. Donada et al. (2023), Y. Jiang et al. (2024), L. Li & Z. Shao (2022), H. Niu et al. (2020), and V. V. Jadhav et al. (2021). To make a proper comparison, each analysis was restricted to binary systems within the same mass-ratio range as those considered in the respective reference studies. When no specific mass-ratio range was reported, we considered all binary systems in our sample. The results are summarized in Table 1.

In summary, the values obtained for the binary fraction are slightly higher than the results found in the literature. The comparisons, which account for binaries with different mass ratios, show that our methodology effectively represents the binary population, regardless of the selection criteria used. The discrepancies observed between the literature results and those obtained in this work may be associated with the different methods used to identify unresolved binaries.

## 6. Dependence of the Binary Fraction on Cluster Properties

Our analysis is based on a large homogeneous sample of 773 OCs for which individual stellar masses were determined, allowing a robust statistical analysis of how environmental properties and stellar mass influence the binary fraction. In this section, we examine the dependence of the estimated binary fraction on cluster properties and explore the occurrence of multiple systems as a function of stellar mass.

### 6.1. Age and Binary Fraction

Figure 9 shows the distribution of $f_b$ as a function of cluster age. We applied a locally weighted scatterplot (LOWESS) smoothing regression and examined the associated uncertainties. Although a slight increase in the binary fraction with age is observed, the large dispersion in $f_b$ values indicates that the binary fraction does not strongly depend on the evolutionary stage of the cluster. This result is consistent with J. R. Hurley et al. (2007), who showed that, on average, the primordial binary fraction remains nearly constant during cluster evolution, except in the final stages approaching cluster dissolution. A similar behavior is found for high mass-ratio binaries.

We find that the mass-ratio distributions of binary systems in young and old clusters, shown in Figure 10, are statistically distinct (adopting a significance level of 0.05) based on the Kolmogorov–Smirnov (KS) test ($p$-value $\sim 5 \times 10^{-66}$) and Anderson–Darling (AD) test ($p$-value = 0.001), while the Mann–Whitney (MW) $U$ test does not indicate a significant difference ($p$-value = 0.27). Similar results were reported by A. C. Childs & A. M. Geller (2025).

For both young and older clusters, we observe a greater preference for high mass ratios, with a peak at $q \approx 0.9$. However, in older clusters, the distribution for $q > 0.6$ shows less variation, while for younger clusters, the peak at $q \approx 0.9$ is much more pronounced, in addition to a considerable decrease in the occurrence of twin binaries. For mass ratios lower than 0.6, the behavior of the two distributions is quite similar, but we can note that older clusters present less variation overall. This difference is possibly a consequence of dynamical interactions that occur throughout the evolution of clusters. There is a greater tendency for wide binaries with low-mass ratios to be disrupted (D. C. Heggie 1975; A. M. Geller et al. 2013; N. R. Deacon & A. L. Kraus 2020), while twin binaries are generally more stable (D. C. Heggie 1975), facilitating their survival over time.

### 6.2. Metallicity and Binary Fraction

Several works in the literature address how metallicity is associated with the formation of multiple systems. For example, both close binaries (M. Moe et al. 2019) and wide binaries (K. El-Badry & H. W. Rix 2019; H.-C. Hwang et al. 2021) show an anticorrelation between the occurrence of these systems and metallicity. A similar result was also found for interacting binaries (V. M. Douna et al. 2015; S. Ponnada et al. 2019). Star formation simulations performed by M. N. Machida et al. (2009) found a greater susceptibility to fragmentation of molecular clouds in metal-poor regions compared to environments with high metallicities. Since most binary systems are likely to form during the fragmentation of molecular clouds (R. L. Kuruwita & T. Haugbølle 2023), the formation of multiple systems is theoretically favored in metal-

**Table 1**
Comparison between the Binary Fractions Estimated in This Work and Those Reported in the Literature

| Common OCs | This Work | | Literature | | | |
|---|---|---|---|---|---|---|
| | $f_b$ Range | Median $f_b$ | $f_b$ Range | Median $f_b$ | $q$ Range | References |
| 114 | 0.17–0.53 | 0.31 | 0.06–0.80 | 0.17 | $q > 0.6$ | J. Donada et al. (2023) |
| 7 | 0.18–0.29 | 0.26 | 0.17–0.22 | 0.19 | $q > 0.6$ | V. V. Jadhav et al. (2021) |
| 17 | 0.22–0.60 | 0.38 | 0.12–0.21 | 0.16 | $q \geqslant 0.5$ | Y. Jiang et al. (2024) |
| 5 | 0.30–0.50 | 0.40 | 0.30–0.46 | 0.41 | $q > 0.2$ | L. Li & Z. Shao (2022) |
| 9 | 0.28–0.55 | 0.40 | 0.29–0.55 | 0.42 | ⋯ | H. Niu et al. (2020) |

**Note.** Only OCs in common with our sample and systems with the same mass ratio selection criteria were considered.

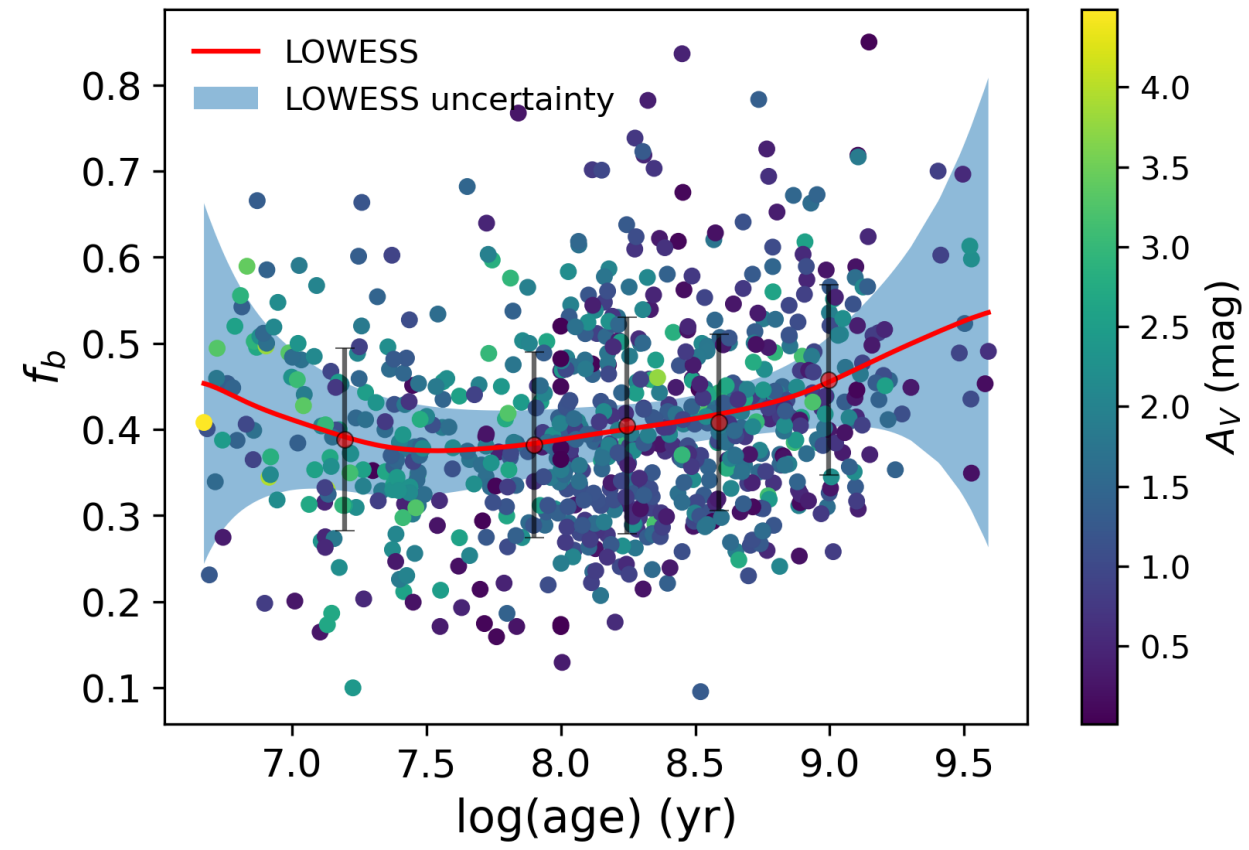

**Figure 9.** Distribution of $f_b$ as a function of cluster age. Points are color-coded by extinction $A_V$. Error bars represent the standard deviation within each bin, with each bin containing approximately 20% of the full sample.

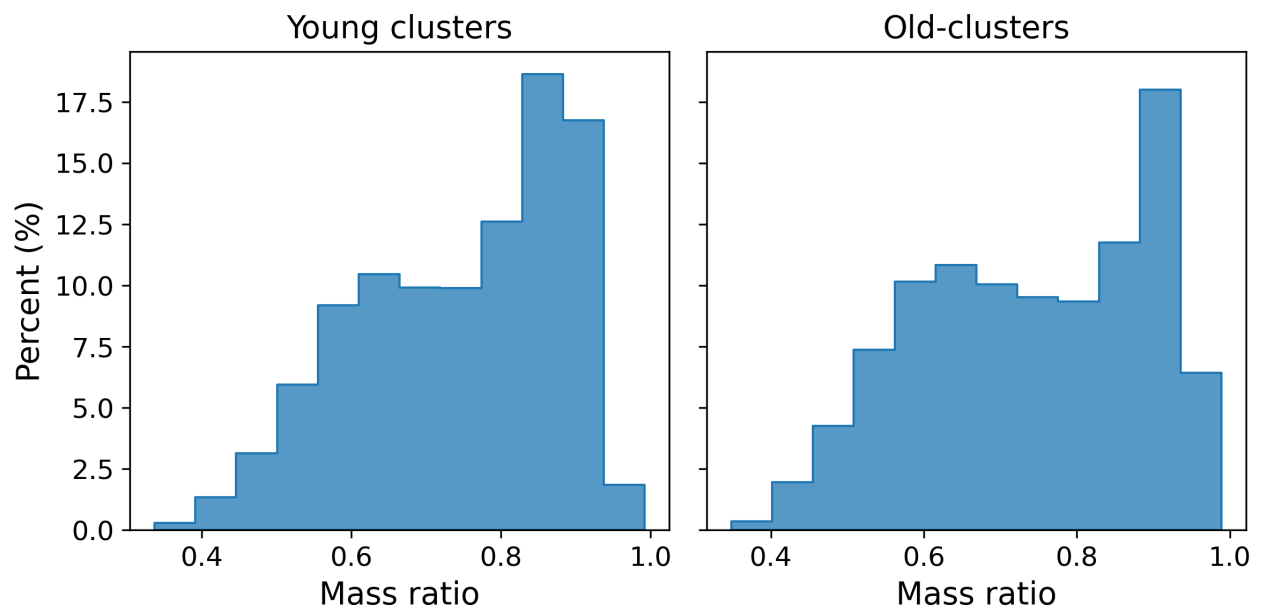

**Figure 10.** Mass-ratio distribution. The left panel shows young OCs (log(age) < 8.0), while the right panel shows old clusters (log(age) > 8.0).

poor clouds. In C. Badenes et al. (2018), it was found that in low-metallicity regions ([Fe/H] $\leqslant -0.5$), the multiplicity fraction is about 2–3 times higher than in metal-rich regions ([Fe/H] $\geqslant 0$).

In the present study, we find a weak anticorrelation between $f_b$ and metallicity. However, this result may be affected by both sample-related biases and the methodology adopted to estimate $f_b$. In particular, Figure 11 shows that the number of clusters at both low and high metallicity is relatively small, leading to larger uncertainties in these regimes. In addition, the correction applied to the binary fraction depends on the properties of the benchmark subsample, which contains relatively few metal-poor clusters (see Figure 16 in Appendix A). These factors should be considered when interpreting the observed anticorrelation. Despite these potential biases, the persistence of the anticorrelation is consistent with theoretical expectations and previous observational results, supporting the idea that lower metallicity environments favor the formation of multiple systems.

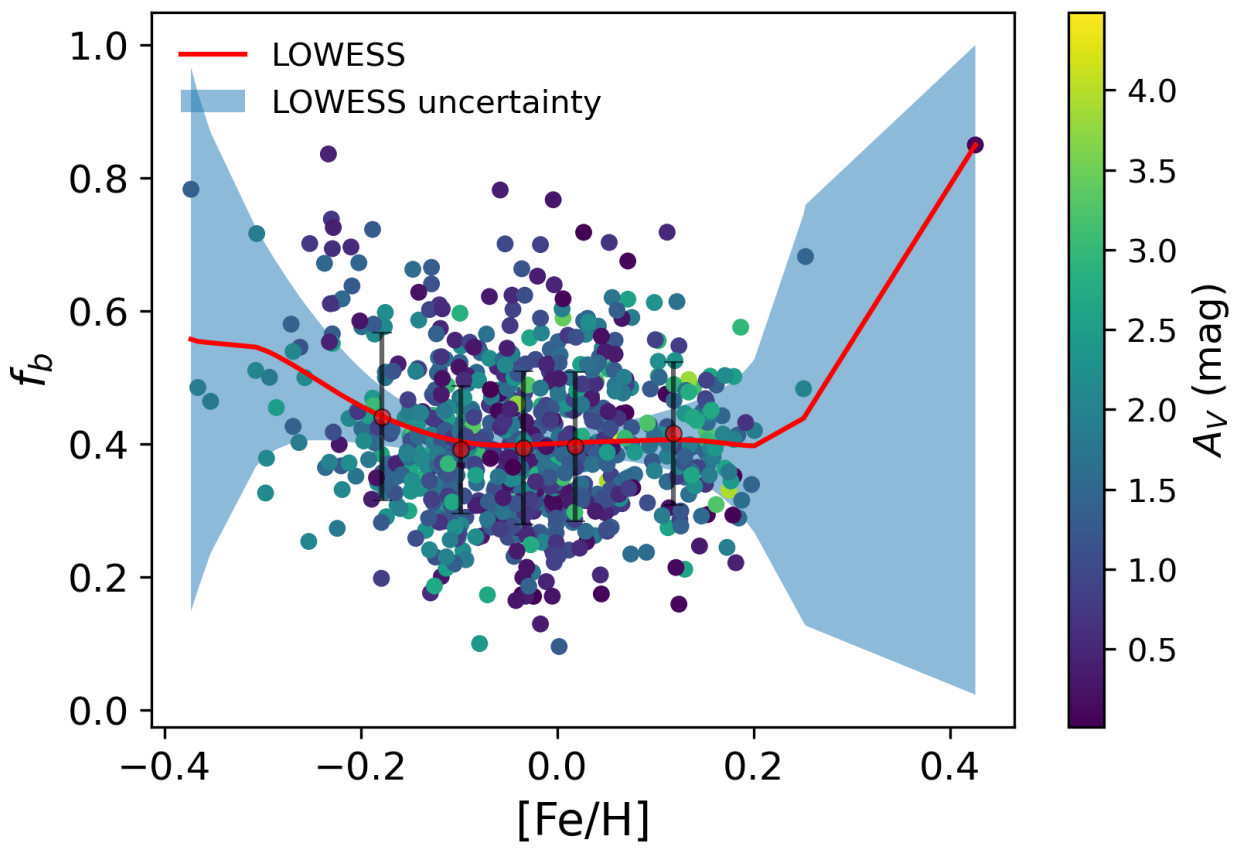

**Figure 11.** Distribution of $f_b$ as a function of cluster metallicity. Points are color-coded by extinction $A_V$.

Furthermore, although the mass ratio distributions in metal-poor ([Fe/H] < 0) and metal-rich ([Fe/H] > 0) clusters are distinct according to the KS, AD, and MW tests, the magnitude of this difference is small. This is illustrated in the cumulative distributions (Figure 12), where the two curves overlap across the entire range of $q$, with only small deviations between $q$ =0 .7 and $q = 0.9$. This suggests that the differences must be purely statistical and may not represent a real physical effect.

### 6.3. Stellar Mass and Binary Fraction

To analyze how stars of different mass ranges behave in binary systems, we used the definitions described in Section 3.5 for $P_{\rm Prim}$, $P_{\rm Comp}$, and $P_{\rm BS}$ in bins of different mass ranges. Mass bins were defined in logarithmic space, with a fixed width of $\Delta \log M = 0.05$ below $4\,M_\odot$ and a variable width above this limit corresponding to linear intervals of $1\,M_\odot$, with each bin containing at least 1000 stars. Figure 13 shows the distributions of $P_{\rm Prim}$, $P_{\rm Comp}$, and $P_{\rm BS}$ as a function of the median stellar mass of each bin.

For low-mass stars, the median value for $P_{\rm BS}$ was about 80%. This high multiplicity fraction in this mass regime is particularly dominated by companion roles, with approximately 70% found as secondary and 12% being primary in a binary system. For intermediate-mass stars, the median values for $P_{\rm Prim}$, $P_{\rm Comp}$, and $P_{\rm BS}$ are 52%, 18%, and 70%, respectively. Additionally, in this mass range, we find that $P_{\rm Prim}$ increases with increasing mass, especially for $M > 4\,M_\odot$. In the high-mass regime, on average, approximately 80% of stars participate in a binary system, values consistent with results found in the literature (H. Sana & C. J. Evans 2011; R. Chini et al. 2012, 2013), with 65% acting as primaries and only 14% being found as secondaries.

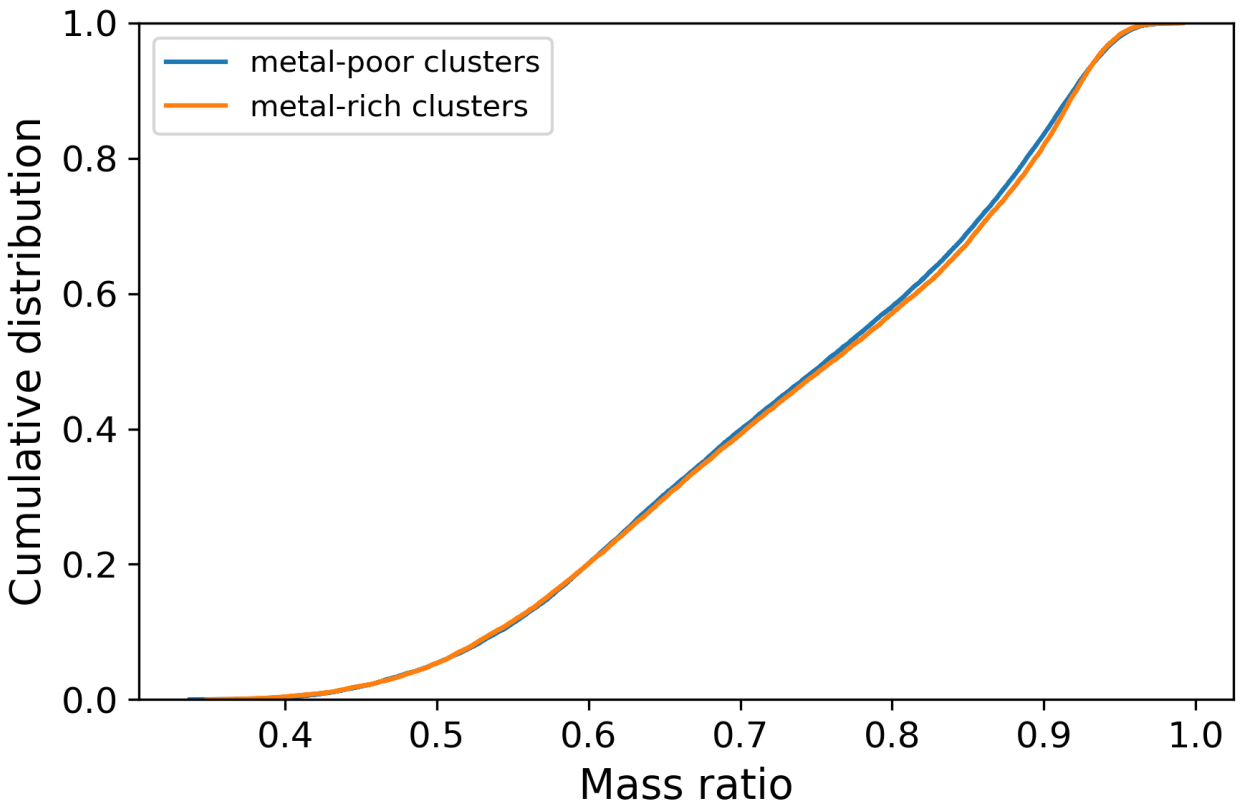

**Figure 12.** Cumulative distribution functions of the mass ratio for binary systems in metal-poor ([Fe/H] < 0; blue) and metal-rich ([Fe/H] > 0; orange) OCs.

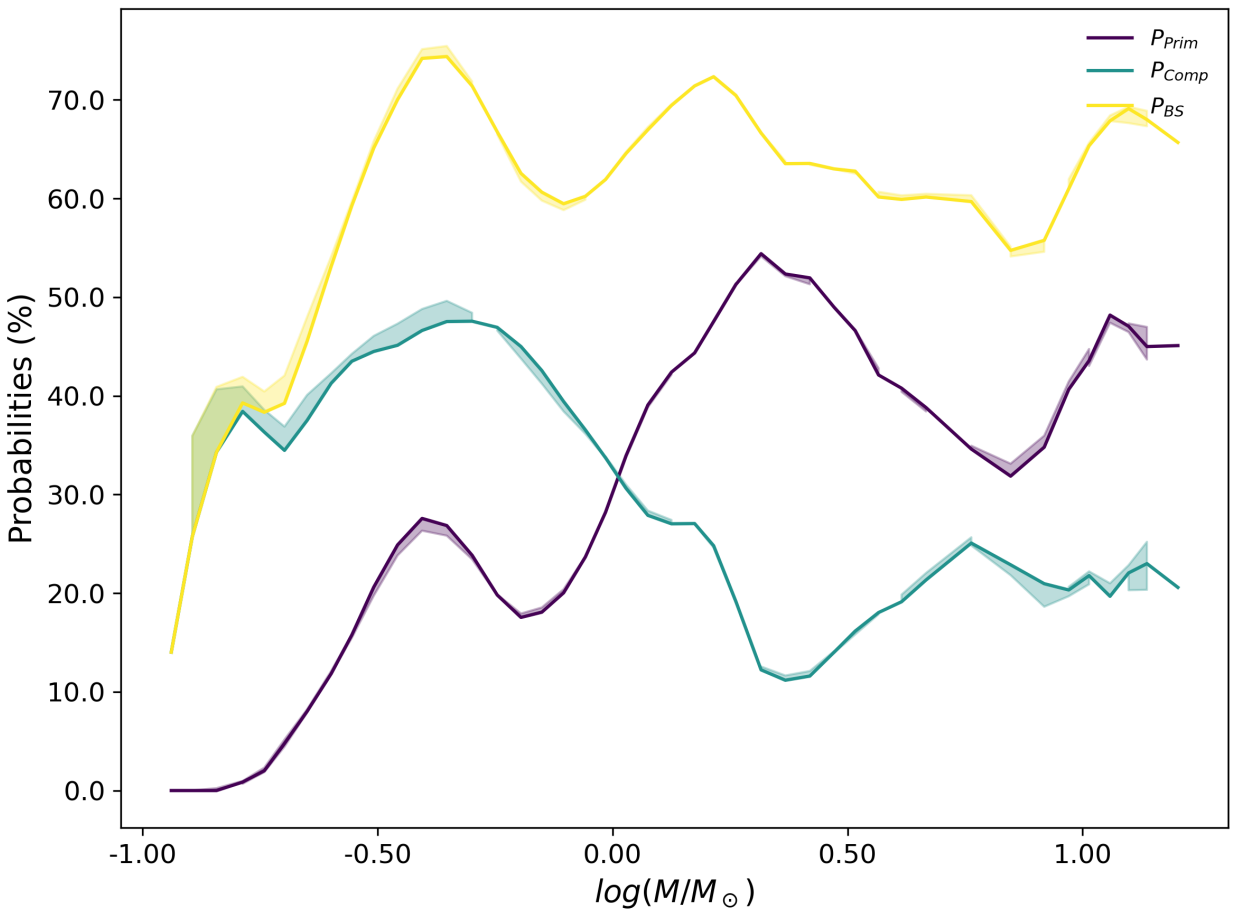

**Figure 13.** Probability that stars within a given mass range belong to binary systems (yellow), and the probabilities of being the primary (purple) or secondary (blue) component.

Furthermore, the results also suggest that with the exception of stars with $M < 0.5\,M_\odot$ or $M > 10\,M_\odot$, the occurrence in multiple systems does not vary significantly with mass, with the value of $P_{\rm BS}$ being around (70 ± 10)%. The tendency for $P_{\rm Prim}$ to increase and $P_{\rm Comp}$ to decrease with stellar mass is expected as the higher gravitational potential of massive stars facilitates the capture and retention of companions. The distinct behavior observed for intermediate-mass stars may reflect the fact that massive stars have a preference for forming binaries with companions with masses around ∼5–8 $M_\odot$, which is the range in which $P_{\rm Prim}$ decays most sharply with mass.

### *6.4. Dynamical Evolution and Binary System Spatial Distributions*

The phenomenon of mass segregation has been observed in several OCs (e.g., I. A. Bonnell & M. B. Davies 1998; C. Zwicker et al. 2024; M. S. Angelo et al. 2025; R. Liu et al. 2025). Due to the dynamic evolution of OCs, high-mass stars tend to concentrate in the central regions, while less massive ones move to the outer regions, which can result in the cluster losing mass over time due to the evaporation of low-mass members. In this context, binary systems play an important role in maintaining the primordial structure of OCs (M. B. N. Kouwenhoven & R. de Grijs 2008; S. Rastello et al. 2020) as binary systems are generally more massive than individual systems and are therefore less likely to escape from the cluster (K. Takahashi & S. F. Portegies Zwart 2000; S. F. Portegies Zwart et al. 2001).

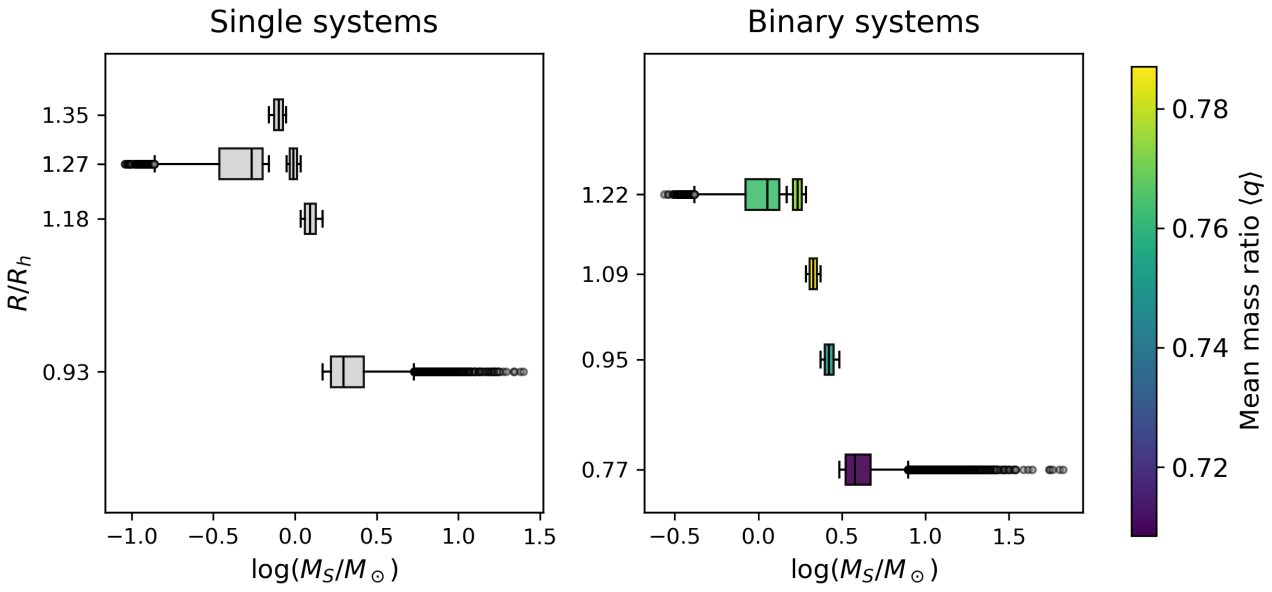

**Figure 14.** Normalized radial distance ($R/R_h$) as a function of the average system mass for all single (left panel) and binary (right) systems in our sample. The color scale represents the average mass ratio of the binary systems.

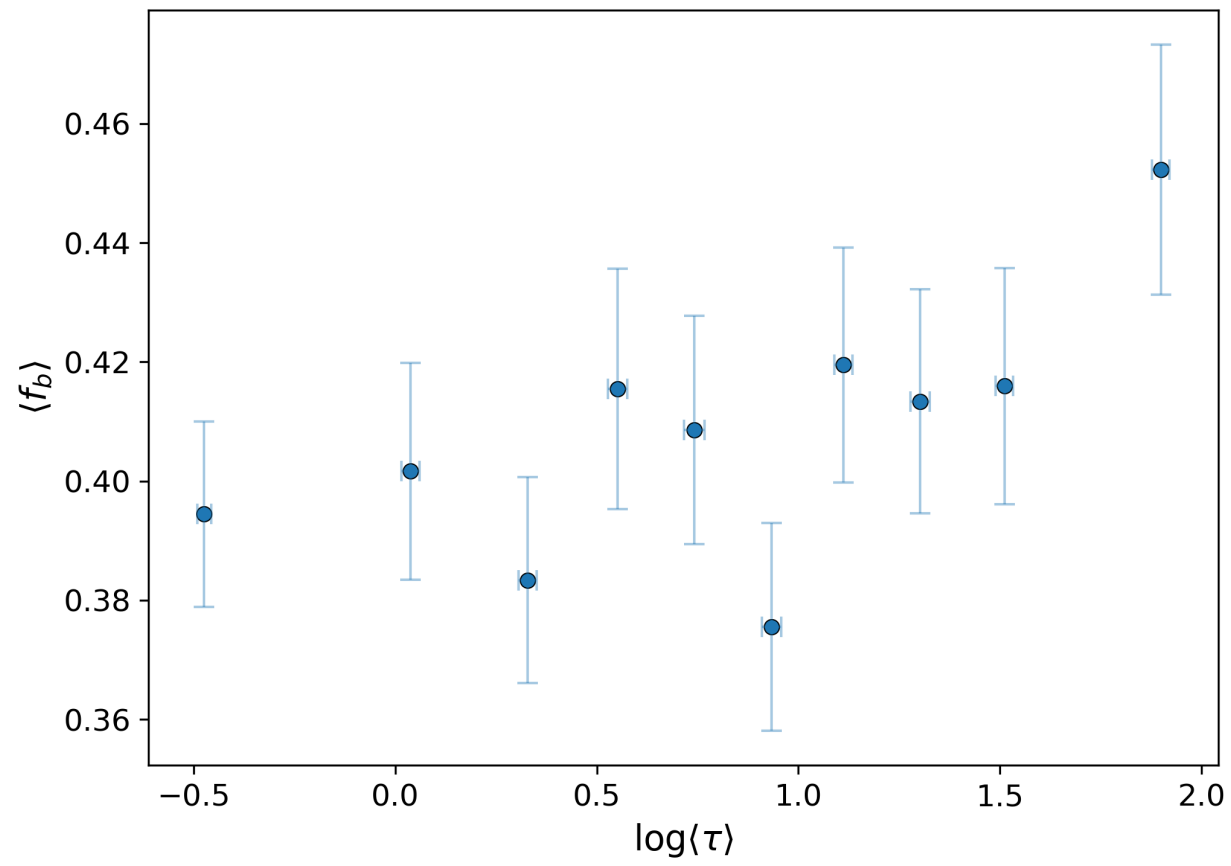

**Figure 15.** Average binary fraction as a function of the cluster's dynamical state parameter $\tau$.

Figure 14 presents the distribution of the systems masses, defined as $M_S = M_1(1 + q)$, as a function of their normalized radial distance ($R/R_h$), both for the single systems (left panel) and for binary systems (right panel). The median $R/R_h$ values obtained for the single and binary systems were 1.20 and 1.04, respectively. Each subset of single stars contains approximately 14,000 systems, while each subset of binaries contains approximately 13,000 systems.

As shown in Figure 14, a clear trend is observed, with the mass of systems increasing toward the more central regions of the clusters, both for binary systems and for single stars. Furthermore, the average binary mass ratio decreases with increasing distance from the cluster center, according to what was observed by M. D. Albrow (2024) in the Hyades and Praesepe clusters.

Figure 15 shows the distributions of the average binary fraction as a function of the parameter $\tau$. An increasing trend of $f_b$ with cluster dynamical state is observed, although the magnitude of this variation is small, rising from 0.40 in nonrelaxed clusters to 0.46 in relaxed clusters. This suggests that, at least on average, dynamic evolution does not substantially modify the binary fraction established during the initial stages of cluster evolution.

## 7. Conclusion

In this study we present a statistical investigation of stellar multiplicity in a sample of 773 OCs, where the binary systems were determined by comparing the observed data with synthetic clusters generated via the Monte Carlo method. For the calculations of the binary fraction, systems with relative uncertainties in companion mass greater than 0.5 were excluded, and as a result the total binary fraction in this study only considers systems with $q \geqslant 0.34$. To mitigate observational biases associated with distance and reddening, we assumed that the binary fraction does not vary significantly with these parameters and adopted a mean-based correction method, using clusters with distances <1.5 kpc and $A_V < 0.5$ mag as a benchmark sample.

We found the total binary fraction ranges from 0.10 to 0.85, with values for high mass-ratio systems ($q > 0.5$) ranging from 0.06 to 0.79. We found a large dispersion of $f_b$ as a function of cluster ages, although we observed a tendency for $f_b$ to increase in older clusters. Similarly, dynamically evolved clusters also showed a higher binary fraction, but the magnitude of this difference is less than 10% when compared to nonrelaxed clusters, suggesting that $f_b$ does not vary strongly with time. Nevertheless, the distinct mass-ratio distributions observed in young and old clusters indicate that dynamical evolution may redistribute binary components, potentially favoring systems with higher mass ratios. We also find an anticorrelation between binary fraction and metallicity, in agreement with previous studies (e.g., C. Badenes et al. 2018; M. Moe et al. 2019; J. Donada et al. 2023). Although a statistical difference is detected between the mass-ratio distributions of metal-poor and metal-rich clusters, its magnitude is small and unlikely to reflect a physical effect.

Furthermore we find that the fraction of stars acting as primary components in binary systems increases with stellar mass, with $\langle P_{\rm prim} \rangle$ rising from ~15% for low-mass stars to ~44% for high-mass stars. Conversely, the probability of a star being a companion decreases with mass, from ~40% at low masses to ~20% at high masses. The overall probability of a star belonging to a binary system remains approximately constant at ~60% over most of the mass range, except for stars with $M < 0.5\,M_\odot$, for which it increases significantly with mass. In addition, we find that systems with primary masses below $1\,M_\odot$ and above $5\,M_\odot$ exhibit similar average mass ratios, $\langle q \rangle \approx 0.75$. However, a pronounced minimum is observed within this interval, reaching $q \approx 0.64$ at $M_1 = 2.75\,M_\odot$.

Finally, statistically, our analysis indicates that binary systems are, on average, more centrally concentrated than single stars, making them less likely to escape from the cluster. This helps to preserve the cluster's structure and primordial mass over time. Furthermore, the average mass ratio decreases as one moves away from the center of the cluster.

### Acknowledgments

R.P.A. is maintained with grants from CAPES. W.S.D. would like to thank CNPq grant 310765/2020-0. H.M. would like to thank FAPEMIG grants APQ-02030-10 and CEX-PPM-00235-12. This research was performed using the facilities of the Laboratório de Astrofísica Computacional da Universidade Federal de Itajubá (LAC-UNIFEI). The LAC-UNIFEI is maintained with grants from CAPES, CNPq, and FAPEMIG. This work has made use of data from the European Space Agency (ESA) mission Gaia (http://www.cosmos.esa.int/Gaia), processed by the Gaia Data Processing and Analysis Consortium (DPAC; http://www.cosmos.esa.int/web/Gaia/dpac/consortium).

## Appendix A
## Benchmark Subsample

The benchmark clusters used in this work are identified in Table 3 through a dedicated flag. This subsample consists of clusters with $A_V < 0.5$ and distances smaller than 1.5 kpc.

In Figure 16, we compare the distributions of the fundamental cluster parameters for the full sample and the benchmark subsample. To quantify these differences, we performed KS, MW $U$, and AD tests. Differences between distributions were considered statistically significant when $p < 0.05$.

For cluster age, all tests indicate no statistically significant difference between the two samples, suggesting that the benchmark clusters span a similar age range as the full sample. For metallicity, the distributions are similar only in the MW test evaluation, while the distributions of the number of members are similar only in the KS test, and the total mass distribution of the clusters are distinct under all tests. The $p$-values for each test are listed in Table 2.

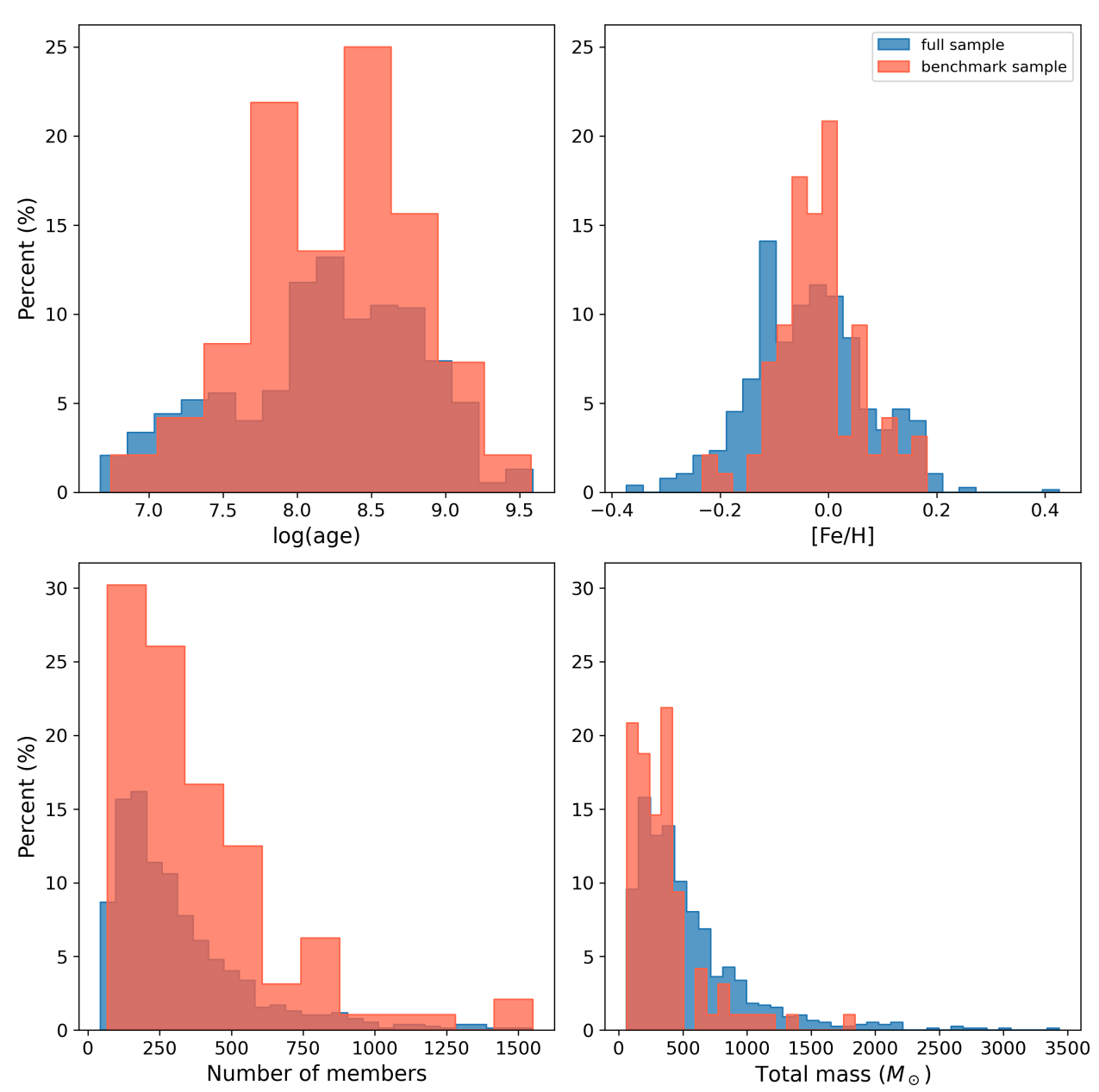

**Figure 16.** Distribution of the parameters of the full sample (blue) and benchmark sample (red).

**Table 2**
Statistical Comparison between the Full and Benchmark Samples

| Parameter | KS ($p$-value) | MW ($p$-value) | AD ($p$-value) |
|---|---|---|---|
| Age | 0.50 | 0.63 | 0.25 |
| [Fe/H] | 0.00 | 0.08 | 0.00 |
| $N_{\text{members}}$ | 0.06 | 0.02 | 0.03 |
| $M_{\text{total}}$ | 0.00 | 0.00 | 0.00 |

**Note.** Columns report the $p$-values of the KS, MW, and AD tests comparing the full and benchmark samples.

## Appendix B
## Results Table

**Table 3**
Binary Fraction (Total and for Systems with $q > 0.5$), the Half-mass Radius, the Relaxation Time, and the Dynamic Evolution Parameter of the Studied Clusters

| Cluster | $f_b$ | $f_{b_{q>0.5}}$ | $R_h$ (pc) | $t_{\text{relax}}$ (Myr) | $\tau$ | Benchmark Flag |
|---|---|---|---|---|---|---|
| ASCC 10 | 0.30 ± 0.15 | 0.26 ± 0.14 | 4.62 ± 0.81 | 51.50 ± 14.52 | 3.80 ± 1.32 | False |
| ASCC 105 | 0.70 ± 0.26 | 0.67 ± 0.25 | 4.37 ± 0.60 | 71.97 ± 16.34 | 1.82 ± 0.47 | False |
| ASCC 107 | 0.68 ± 0.26 | 0.65 ± 0.25 | 2.22 ± 0.28 | 14.92 ± 3.15 | 3.01 ± 2.57 | False |
| ⋮ | | | | | | |
| vdBergh 130 | 0.39 ± 0.14 | 0.37 ± 0.15 | 1.31 ± 0.13 | 6.09 ± 1.09 | 0.91 ± 0.31 | False |
| vdBergh 80 | 0.52 ± 0.18 | 0.49 ± 0.16 | 2.24 ± 0.38 | 19.75 ± 5.35 | 0.43 ± 0.17 | False |
| vdBergh 92 | 0.41 ± 0.09 | 0.40 ± 0.09 | 2.47 ± 0.25 | 36.94 ± 6.72 | 0.18 ± 0.05 | False |

**Note.** The full version of this table containing the results of all clusters is provided online.

(This table is available in its entirety in machine-readable form in the online article.)

## ORCID iDs

Ruan P. Alves https://orcid.org/0009-0001-8769-0512
Hektor Monteiro https://orcid.org/0000-0002-0596-9115
Wilton S Dias https://orcid.org/0000-0002-5443-9524
Gabriel R. Hickel https://orcid.org/0009-0002-0280-9877